\documentclass[prd,superscriptaddress,onecolumn,showpacs,11pt,msmath,
preprintnumbers,showkeys]{revtex4}
\usepackage{wrapfig,rotating}
\usepackage{graphicx,epsfig,color}



\usepackage{graphicx}
\usepackage{dcolumn}
\usepackage{bm}

\def\beq{\begin{equation}}
\def\eeq{\end{equation}}
\def\beeq{\begin{eqnarray}}
\def\eeeq{\end{eqnarray}}

\begin{document}
\title{Hot Spot model  of Nucleon  and Double Parton Scattering.}

\author{B.\ Blok$^{1}$, R. Segev$^{1}$,
 M.\ Strikman$^{2}$
\\[2mm] \normalsize $^1$ Department of Physics, Technion -- Israel Institute of Technology,
Haifa, Israel\\
\normalsize $^2$ Physics Department, Penn State University, University Park, PA, USA}

\begin{abstract}
\par  We calculate the
rate of 
double parton scattering (DPS) in proton - proton collisions in the framework of the recently proposed hot spot model of the nucleon structure.  The resulting rate, especially for the case of three hot spots, appears to be in tension with the current experimental data 
on DPS at the LHC.

\end{abstract}
\maketitle
\thispagestyle{empty}

\vfill
\section{Introduction}
\par The 3D structure of nucleons  has been attracting attention  at least since discovery of quarks.
For a single parton distributions factorization theorems have allowed to investigate longitudinal momentum plus transverse coordinate single parton distributions (Generalized parton distributions)
 This is one of the central topics that will be studied in the future EIC collider to be built at Brookhaven National Laboratory
\cite{EIC}.

Probing correlations between the partons requires  more complicated tools like four jet production, for a review see for example Ref. \cite{BS}.

The nonperturbative correlations were considered in the constituent quark model to explain the success of the additive quark model, for a review see  Ref.  \cite{levin} . Small size (hot spot)   correlations generated  by the QCD evolution were introduced in Ref. \cite{mueller1}. Recently the multi hot spot model of nucleon was introduced  in \cite{hs1,Mantysaari:2022ffw,Mantysaari:2022sux}.  The parameters of the model were fixed  by fitting the  cross section  of reaction $\gamma p \to J/\psi + gap + Y$ within the model \cite{hs1}  which  assumes that 
fluctuations of the gluon field at a wide range of momentum transfer satisfy  the Good Walker  relation \cite{GW}, for  a recent review of conditions of applicability of the Good - Walker model see \cite{progressinphsyics}.

\par In the last decade  
 a lot of progress, both theoretical \cite{TP,M,16a,16b,16c,16d,16e,16g,16k,16l,16n} and experimental  \cite{ATLAS,CMS}
 has been made  in our understanding  of the Double-Parton Scattering (DPS) which are sensitive to parton - parton correlations in transverse (relative to the. hadron high momentum) plan. 

 The DPS cross section 
is usually characterized by the so called effective cross section defined as
 \beq
\sigma_{DPS}=\frac{\sigma_{1}\sigma_{2}}{\sigma_{eff}},
\eeq
 where $\sigma_{DPS}$ is the cross-section of the DPS process, $\sigma_{1}$ and $\sigma_{2}$ are the cross-sections of the individual hard partonic interactions, while  $\sigma_{eff}$ depends heavily on the inner structure of the colliding hadrons.
\par In this work we will use the hot-spot model with the  parameters  found  in Ref. \cite{Mantysaari:2022ffw,Mantysaari:2022sux} to calculate $\sigma_{eff}$ .
\par We  demonstrate that  $\sigma_{eff}$
 strongly depends on the parameters of the hot spot model.
The authors of  \cite{Mantysaari:2022ffw,Mantysaari:2022sux}  identify two sets of parameters compatible with DIS and their model of rapidity gap processes 
for $N_q=3$ and $N_q=7$ hot spots respectively.
\par For the set with variable $N_{q}=7$ we find $\sigma_{eff}\approx 17$ mb, and for the set with $N_{q}\equiv3$ we get $\sigma_{eff}\approx 10.5$ mb.
\par The experimental data for $\sigma_{eff}$ are  $\approx 20$  mb. This experimental data are
 however available
at moderate  values of $Q\sim 20$ GeV and higher. The inverse evolution using DGLAP along the lines of \cite{16e,16g,16k} leads to $\sigma_{eff}$ of order 25-35
mb at low scales of several GeV where hot spot model is usually formulated. We see the tension between experimental data on 
DPS and the DPS cross section calculated in the  hot spot model, especially for $N_q=3$ case, which is substantially higher than the experimental one.
\par This paper is organized as follows. Section 2  we review  the mean-field approach to MPI and in section 3 we review  the details of the hot spots model. In section  4  we calculate the effective cross-section using the hot spots model. In section IV we compare our results with measurements to find the limits of this model and  present  our conclusions.

\section{The Mean-Field Approach To MPI}
\par The hot spot model  is
formulated in
the region of relatively small $Q^2$, where one can neglect the DGLAP evolution.
Hence we can use the parton model to calculate the DPS cross sections.
\par Recall that in the parton model approach the DPS cross section is expressed through convolution of  two particle
Generalized Parton Distributions $_2GPD$ s \cite{16b}.
\beq
\frac{1}{\sigma_{eff} }=\frac{\int \frac{d^2\Delta}{(2\pi)^2}_2G(x_1,x_2,Q_1^2,Q_2^2,\Delta)+_2G(x_3,x_4,Q_1^2,Q_2^2,\Delta)}{f(x_1,Q_1^2)f(x_2,Q_2^2)f(x_3,Q_3^2)f(x_4,Q_4^2)}.
\label{2}
\eeq
Here $\vec \Delta$ is the momentum conjugate to the
transverse
 distance between two partons participating in the DPS process
(see Fig.1).
\begin{figure}[htbp]
\begin{center}
\includegraphics[scale=0.45]{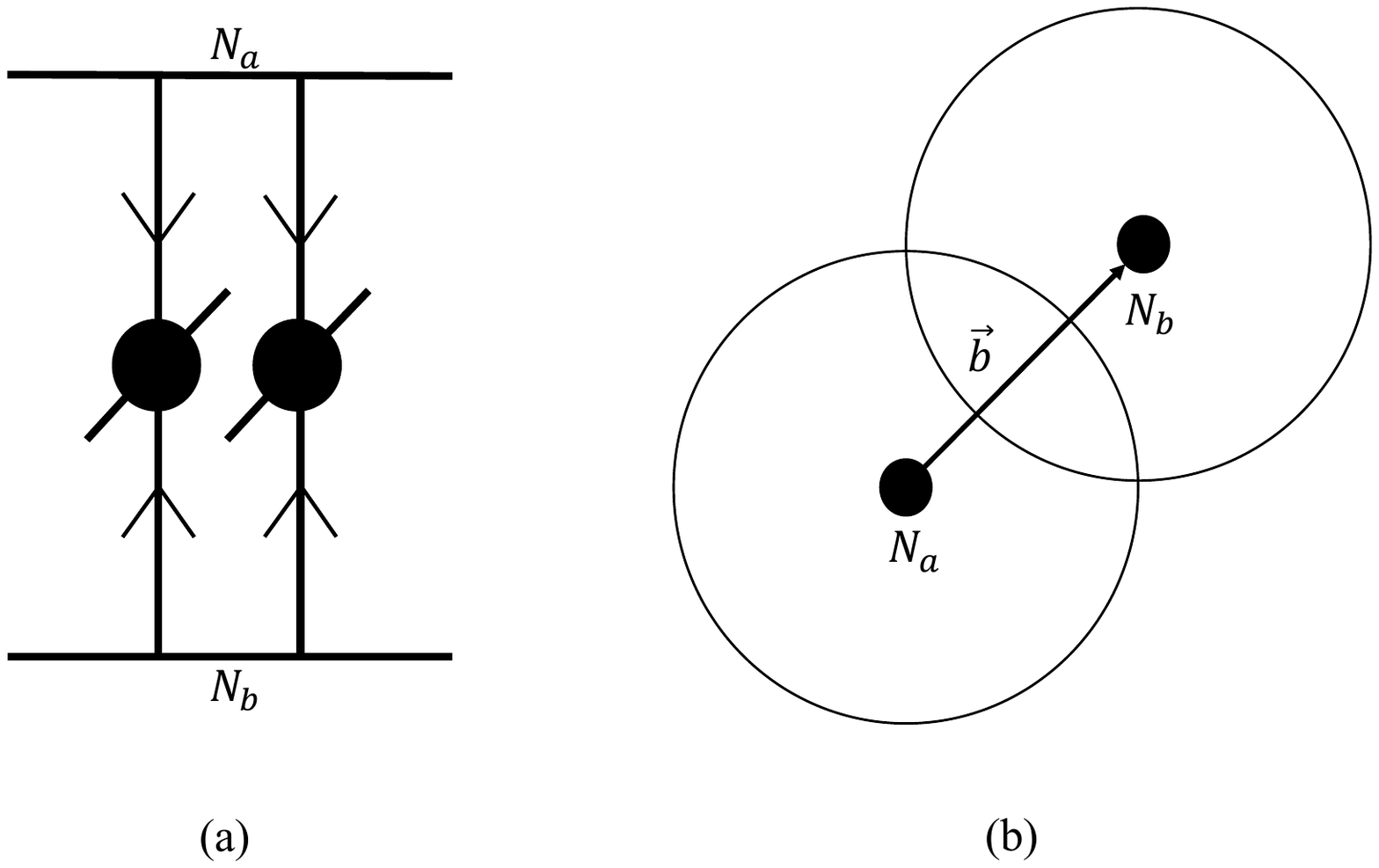}
\caption{.Fig.1a: Parton model contribution to Double Parton Scattering of two nucleons ;Fig.1b Collision of two nucleons 
at the
 impact parameter b}
\end{center}
\end{figure}

In the mean field approximation, that is valid at small transverse scales of order several GeV and small x \cite{16b}, one can prove that the two particle GPDs factorize:
\beq
_2G(x_1,x_2,Q_1^2,Q_2^2,\Delta)=_1G(x_1,Q_1^2,\Delta)_1G(x_1,Q_1^2,\Delta),
\eeq
where $_1G$ are the conventional one particle GPD \cite{diehl,radyushkin}. The latter in the mean field approximation can be written as 
\beq
_1G(x_1,Q_1^2,\Delta)=f(x_1,Q_1^2)F_{2g}(\Delta,x_1),
\eeq
where $F_{2g}$ is called a two gluon formfactor and only weakly depends on x and $Q^2$.
\par In the coordinate space we have
\beq
_1G(x,Q^2,\vec r)=f(x,Q^2)\rho(\vec r); \rho(\vec r)=\int d^2\Delta F_{2g}(\Delta,x_1)\exp(i\vec \Delta\vec r),
\eeq
where $\rho(\vec r)$ is the transverse parton density. 
Note that in such approach the parton density is normalized by one
\beq
\int \rho(\vec r)d^2r=1.
\label{norm1}
\eeq
The effective cross section is then given by
\beq
\frac{1}{\sigma_{eff}}=\int d^2b(\int d^2r\rho(\vec r)\rho(\vec r-\vec b))^2,
\label{3}
\eeq
where $\vec b$ is the impact parameter of the proton proton collision.

\section{The Hot Spots Model.}
\par The hot spots mode
assumes  a specific type of distribution for the transverse positions 
of the gluonic content of the proton \cite{Mantysaari:2022ffw,Mantysaari:2022sux}. According to the model the gluons are concentrated around $N_q$ points, called the hot spots, positioned in the transverse positions 
$\vec{b}_i$ with a two-dimensional Gaussian distribution around the center of mass the proton, marked as $\vec{c}$, with the width  $B_p$. The hot spots distribution around a known center is :
\beq
\rho\left(\left\{ \vec{b}_{i}\right\} _{i=1}^{N_{q}}|\vec{c}\right)=\frac{2\pi B_{p}}{N_{q}}\delta^{\left(2\right)}\left(\frac{\sum_{i=1}^{N_{q}}\vec{b}_{i}}{N_{q}}-\vec{c}\right)\left(\prod_{i=1}^{N_{q}}\frac{1}{2\pi B_{p}}e^{-\frac{\left(\vec{b}_{i}-\vec{c}\right)^{2}}{2B_{p}}}\right),
\eeq
 where the normalization factor of $2\pi B_{p}/N_{q}$ is chosen to get a total integral of one. Each hot spot has the Gaussian density around the center of the hot spot with a width of $B_q$ and can have a  fluctuating strength denote  as $p_i$.
 Hence, the probability distribution   
 to have a hard parton  at  position $\vec{r}$ is given by:
\beq
\rho\left(\vec{r},\left\{\vec{b}_{i},p_{i}\right\}_{i=1}^{N_q}|\vec{c}\right)=\frac{2\pi B_{p}}{N_{q}}\delta^{\left(2\right)}\left(\frac{\sum_{i=1}^{N_{q}}\vec{b}_{i}}{N_{q}}-\vec{c}\right)\left(\prod_{i=1}^{N_{q}}\frac{1}{2\pi B_{p}}e^{-\frac{b_{i}^{2}}{2B_{p}}}\right)\left(\frac{1}{N_{q}}\sum_{i=1}^{N_{q}}p_{i}\frac{1}{2\pi B_{q}}e^{-\frac{\left(\vec{r}-\vec{b}_{i}\right)^{2}}{2B_{q}}}\right).
\label{start}
\eeq
Here the hot spot strengths $p_i$ are assumed to have random distribution 
\beq
P(\log(p_i))=\frac{1}{\sqrt{2\pi}}\exp(-\log(p_i)^2/2\sigma^2), \label{6}
\eeq
so that $\bar p_i\equiv E(p_i)$ -average value of $p_i$ is equal to $\exp(\sigma^2/2)$, and overall normalisation 
is chosen to ensure normalization condition  \ref{norm1}.
\par Using the distribution \ref{6} we obtain for the average value of $p^n$ $E(p^n)$
\beq
E\left[p^n\right]=e^{\frac{n^2\sigma^2}{2}}.
\label{pton}
\eeq
\par Note that  if we take into account the fluctuating strength in order to satisfy the normalization condition
(\ref{norm1}) we
would have 
 to divide the average density by factor  $E(p)$ and the product of four densities that appears in the formula for the  cross section by 
\beq
E\left[p\right]^4=e^{2\sigma^2}.
\label{pdfassig}
\eeq

\section{Calculating the DPS Effective Cross-Section.}
\par In order to calculate the effective cross section we need to calculate the event by event cross section for given
positions of hot spots and impact parameter $\vec{b}$, and the hot spot strenghts using eq. \ref{3}, and then average over the hot spot
positions, impact parameter and hot  spot strengths.
The average of the hot  spots positions is done by  taking  an integral over the positions of the hot-spots, marked as $\left\{\vec{a}_i,\vec{b}_j\right\}_{i,j=1}^{N_q}$ in addition to the collision impact parameter $\vec{b}$.  Next we shall average over the hot spot strength fluctuations using Eqs. \ref{6},\ref{pton},\ref{pdfassig}.

We start by finding the convolution of the single hot spot collision, obtaining the following integral:
\begin{eqnarray}
\int d^{2}r\rho\left(\vec{r},\left\{ \vec{a}_{i}\right\} _{i=1}^{N_{q}},\vec{c}\right)\rho\left(\vec{r},\left\{ \vec{b}_{j}\right\} _{j=1}^{N_{q}},\vec{c}+\vec{b}\right)&\propto&\sum_{i,j}\int d^{2}re^{-\frac{\left(\vec{r}-\vec{a}_{i}\right)^{2}+\left(\vec{r}-\vec{b}_{j}\right)^{2}}{2B_{q}}}\nonumber\\[10pt]
&=&\pi B_q\sum_{i,j}e^{-\frac{\left(\vec{a}_{i}-\vec{b}_{j}\right)^{2}}{4B_{q}}}.
\end{eqnarray}
\par This integral is proportional to the probability for a single hard partonic process to occur for general positions of the hot spots. Taking the square of this expression and integrating  it over the hot spots positions to find $\left(\sigma_{eff}\right)^{-1}$. To do that we need to separate the sums into three different classes. If the positions of the hot spots are marked as $\vec{a}_{i_1},\;\vec{a}_{i_2},\;\vec{b}_{j_1}$ and $\vec{b}_{j_2}$ we write the classes as (Fig.2):
\begin{figure}[htbp]
\begin{center}
\includegraphics[scale=0.45]{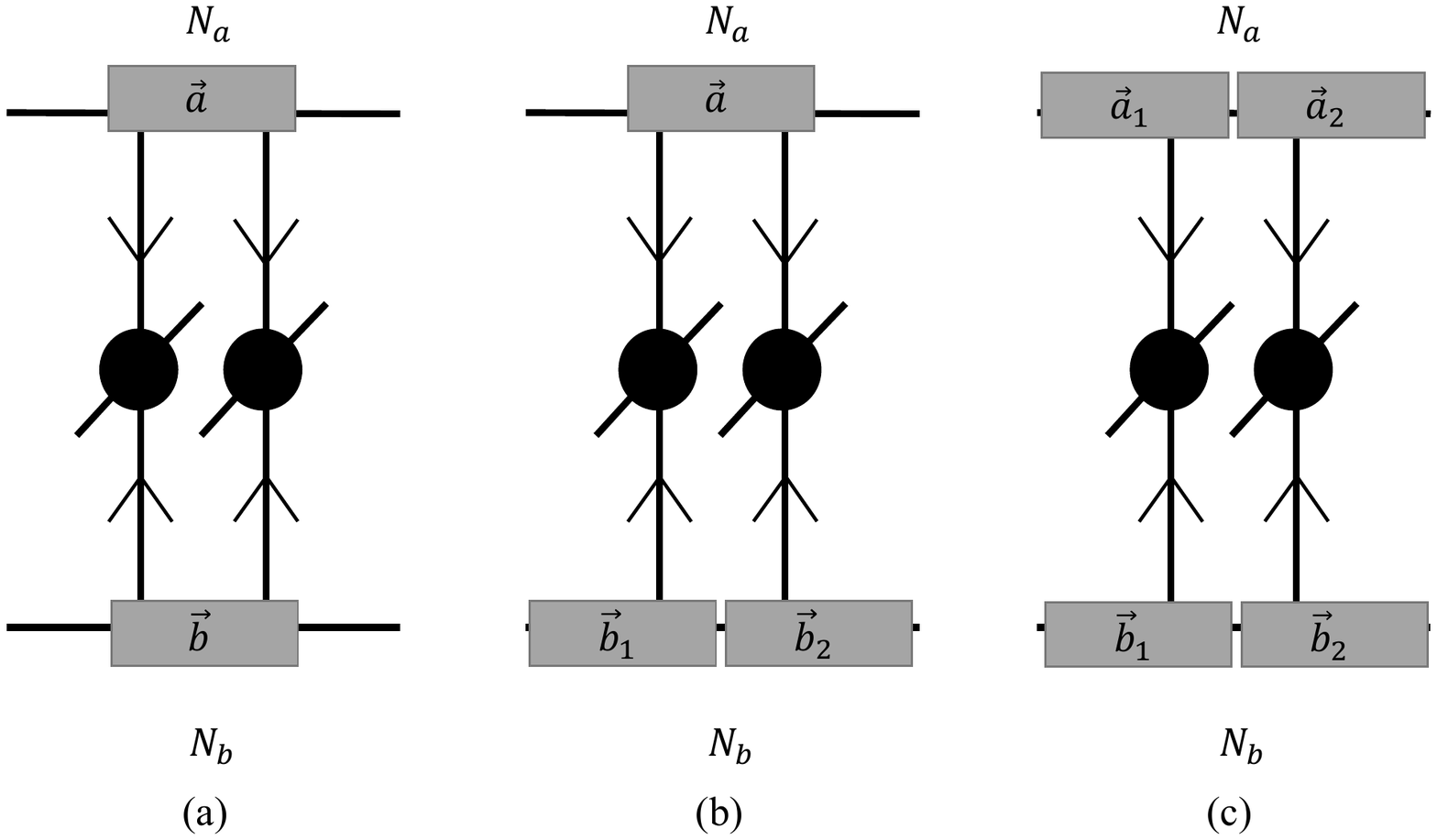}
\label{Fig2a}
\caption{Three distinct classes of diagrams  for DPS scattering in the hot spots model}
\end{center}
\end{figure}
\begin{itemize}
 \item{\bf Class I}: The two partons come from one hot spot for both protons, or $i_1=i_2$ and $j_1=j_2$.
 \item{\bf Class II}: One proton emits the two partons from a single hot spot while the other emits them from two different hot-spots, or $i_1=i_2$ and $j_1\neq j_2$ or $i_1\neq i_2$ and $j_1=j_2$.
 \item{\bf Class III}: Each proton emits the two partons from different hot spots, or  $i_1\neq i_2$ and $j_1\neq j_2$.
\end{itemize}
\par In addition to separating the sums into cases we also use change the delta function to a form more convenient 
in the present calculation:
\beq
\delta^{\left(2\right)}\left(\frac{\sum_{i=1}^{N_{q}}\vec{a}_{i}}{N_{q}}-\vec{c}\right)\delta^{\left(2\right)}\left(\frac{\sum_{j=1}^{N_{q}}\vec{b}_{j}}{N_{q}}-\left(\vec{c}+\vec{b}\right)\right)=\int\frac{d^{2}s_{1}d^{2}s_{2}}{\left(2\pi\right)^{4}}e^{i\vec{s}_{1}\cdot\left(\frac{\sum_{i=1}^{N_{q}}\vec{a}_{i}}{N_{q}}-\vec{c}\right)}e^{i\vec{s}_{2}\cdot\left(\frac{\sum_{j=1}^{N_{q}}\vec{b}_{j}}{N_{q}}-\left(\vec{c}+\vec{b}\right)\right)}
\eeq
\par We also need the integral over a hot spot that isn't part of the collision, it is simply:
\beq
I\left(\vec{c},\vec{s}\right)\equiv\int d^{2}b_je^{\frac{i}{N_{q}}\vec{s}\cdot\vec{b}_j-\frac{\left(\vec{b}_j-\vec{c}\right)^{2}}{2B_{p}}}=2\pi B_{p}e^{-\frac{B_{p}}{2N_{q}^{2}}s^{2}+i\frac{1}{N_{q}}\vec{s}\cdot\vec{c}},
\eeq
\par and the total constant factor is $\left(\left(2\pi\right)^2N_{q}^{2}\left(2\pi B_{p}\right)^{N_{q}-1}\left(2\pi B_{q}\right)\right)^{-2}$, we also set the center of the first proton to be $\vec{c}=\vec{0}$ and the second is then $\vec{c}+\vec{b}=\vec{b}$.
\subsection{Two Partons From a Single hot spot Of Each Proton}
\par In case I the two sums become:
\beq
\left(\sum_{i_{1},j_{1}}p_{i_1}\tilde{p}_{j_1}e^{-\frac{\left(\vec{a}_{i_{1}}-\vec{b}_{j_{1}}\right)^{2}}{4B_{q}}}\right)\left(\sum_{i_{2},j_{2}}p_{i_2}\tilde{p}_{j_2}e^{-\frac{\left(\vec{a}_{i_{2}}-\vec{b}_{j_{2}}\right)^{2}}{4B_{q}}}\right)\rightarrow p^2\tilde{p}^2{N_q}^2e^{-\frac{\left(\vec{a}_1-\vec{b}_1\right)^{2}}{2B_{q}}},
\eeq
\par 
\noindent
where we get a factor of $N_q^2$ from choosing a single hot spot in each proton without loss of generality and $\tilde{p}$ represent the hot spot strength from the $b$ proton. The  $2N_q-2$ hot spots in two nucleons that are  not involved in the interaction give us a factor of $\left(I\left(\vec{0},\vec{s}_1\right)I\left(\vec{b},\vec{s}_2\right)\right)^{N_q-1}$. We are left with the following integral:
\beq
A=\frac{p^2\tilde{p}^2}{4N_{q}^{4}\left(2\pi\right)^{6}B_{q}^{2}}\int d^{2}bd^{2}s_{1}d^{2}s_{2}d^{2}a_1d^{2}b_1e^{-\frac{\left(N_{q}-1\right)B_{p}}{2N_{q}^{2}}\left(s_{1}^{2}+s_{2}^{2}\right)-i\frac{1}{N_{q}}\vec{s}_{2}\cdot\vec{B}-\frac{\left(\vec{a}_1-\vec{b}_1\right)^{2}}{2B_{q}}+\frac{i}{N_{q}}\left[\vec{s}_{1}\cdot\vec{a}_1+\vec{s}_{2}\cdot\vec{b}_1\right]-\frac{a^{2}+\left(\vec{b}_1-\vec{b}\right)^{2}}{2B_{p}}}.
\eeq
\par We are left with a $10$-dimensional Gaussian, but really, the two Cartesian 
coordinates are completely separable so really we can write 
integral $A$  as the square of a $5$-dimensional Gaussian. 
If  we write the parameters as a vector 
 $\vec{x}^T=\left(a_{1x},b_{1x},b_x,s_{1x},s_{2x}\right)$ we obtain:
\beq
A=\frac{p^2\tilde{p}^2}{4N_{q}^{4}\left(2\pi\right)^{6}B_{q}^{2}}\left(\int d^{5}xe^{-\vec{x}^{T}M_A\vec{x}}\right)^{2},
\eeq
\par with $M$ being the following symmetric matrix:
\beq
M_A=\left(\begin{array}{ccccc}
\frac{B_{p}+B_{q}}{2B_{p}B_{q}} & -\frac{1}{2B_{q}} & 0 & -\frac{i}{2N_{q}} & 0\\
-\frac{1}{2B_{q}} & \frac{B_{p}+B_{q}}{2B_{p}B_{q}} & -\frac{1}{2B_{p}} & 0 & -\frac{i}{2N_{q}}\\
0 & -\frac{1}{2B_{p}} & \frac{1}{2B_{p}} & 0 & \frac{i}{2N_{q}}\\
-\frac{i}{2N_{q}} & 0 & 0 & \frac{N_{q}-1}{2N_{q}^{2}}B_{p} & 0\\
0 & -\frac{i}{2N_{q}} & \frac{i}{2N_{q}} & 0 & \frac{N_{q}-1}{2N_{q}^{2}}B_{p}
\end{array}\right),
\eeq
\par and we can use the Gaussian formula $\int d^{n}xe^{-\vec{x}^{T}M\vec{x}}=\frac{\left(2\pi\right)^{n/2}}{\sqrt{\det M}}$ to get:
\beq
A=\frac{p^2\tilde{p}^2}{8\pi B_q {N_q}^2}.
\eeq
\par Averaging over $p$ and $\tilde{p}$ we obtain:
\beq
A=\frac{E\left[p^2\right]^2}{8\pi B_q {N_q}^2}=\frac{e^{4\sigma^2}}{8\pi B_q {N_q}^2}.
\label{caseIfinal}
\eeq
\par This expression takes into account fluctuations of   
the hot spot strength. If we neglect the fluctuations of the hot spot strength,  which corresponds to
 setting $\sigma=0$, we would get:
\beq
A_{\sigma=0}=\frac{1}{8\pi B_q {N_q}^2}.
\eeq
\subsection{Two Partons From a Single Hot Spot Of One Proton And Two Different Hot Spots From The Other Proton}
\par In case II the two sums become one of two sub-cases, for $i_1=i_2$ but $j_1\neq j_2$ we get:
\beq
\left(\sum_{i_{1},j_{1}}p_{i_1}\tilde{p}_{j_1}e^{-\frac{\left(\vec{a}_{i_{1}}-\vec{b}_{j_{1}}\right)^{2}}{4B_{q}}}\right)\left(\sum_{i_{2},j_{2}}p_{i_2}\tilde{p}_{j_2}e^{-\frac{\left(\vec{a}_{i_{2}}-\vec{b}_{j_{2}}\right)^{2}}{4B_{q}}}\right)\rightarrow p^2\tilde{p}_{1}\tilde{p}_{2}{N_q}^2\left(N_q-1\right)e^{-\frac{\left(\vec{a}_1-\vec{b}_{1}\right)^{2}+\left(\vec{a}_1-\vec{b}_{2}\right)^{2}}{4B_{q}}},
\eeq
\par where the factor of ${N_q}^2\left(N_q-1\right)$ comes from choosing the hot spots without loss of generality. In this sub-case, we also get a factor of $\left(I\left(\vec{0},\vec{s}_1\right)\right)^{N_q-1}\left(I\left(\vec{b},\vec{s}_2\right)\right)^{N_q-2}$, leaving us with the integral:
\beq
B_1=\frac{\left(N_{q}-1\right)p^2\tilde{p}_{1}\tilde{p}_{2}}{2N_{q}^{5}\left(2\pi\right)^{7}B_{p}B_{q}^2}\left(\int d^{6}xe^{-\vec{x}^{T}M_{B_1}\vec{x}}\right)^{2},
\eeq
\par where now $\vec{x}^{T}=\left(a_{1x},b_{1x},b_{2x},b_x,s_{1x},s_{2x}\right)$ and:
\beq
M_{B_1}=\left(\begin{array}{cccccc}
\frac{B_{p}+B_{q}}{2B_{p}B_{q}} & -\frac{1}{4B_{q}} & -\frac{1}{4B_{q}} & 0 & -\frac{i}{2N_{q}} & 0\\
-\frac{1}{4B_{q}} & \frac{B_{p}+2B_{q}}{4B_{p}B_{q}} & 0 & -\frac{1}{2B_{p}} & 0 & 0\\
-\frac{1}{4B_{q}} & 0 & \frac{B_{p}+2B_{q}}{4B_{p}B_{q}} & -\frac{1}{2B_{p}} & 0 & -\frac{i}{2N_{q}}\\
0 & -\frac{1}{2B_{p}} & -\frac{1}{2B_{p}} & \frac{1}{B_{p}} & 0 & -\frac{i}{2N_{q}}\\
-\frac{i}{2N_{q}} & 0 & 0 & 0 & \frac{N_{q}-1}{2N_{q}^{2}}B_{p} & \frac{i}{N_{q}}\\
0 & 0 & -\frac{i}{2N_{q}} & -\frac{i}{2N_{q}} & \frac{i}{N_{q}} & \frac{N_{q}-2}{2N_{q}^{2}}B_{p}
\end{array}\right).
\eeq
\par Using the Gaussian formula we get
\beq
B_1=\frac{\left(N_{q}-1\right)p^2\tilde{p}_{1}\tilde{p}_{2}}{4\pi\left(B_p+2B_q\right)N_q^2}.
\eeq
\par For the second sub-case, $i_1\neq i_2$ and $j_1=j_2$, it can be shown that we get the same constant factor but a different matrix, but overall the determinants are the same so we get $B_2=B_1\Rightarrow B=2B_1$. The average over the hot spots strengths  gives  the final expression  for this case:
\beq
B=\frac{\left(N_{q}-1\right)E\left[p^2\right]E\left[p\right]^2}{2\pi\left(B_p+2B_q\right){N_q}^2}=\frac{\left(N_{q}-1\right)e^{3\sigma^2}}{2\pi\left(B_p+2B_q\right){N_q}^2},
\label{caseIIfinal}
\eeq
\par In the case when  fluctuations of the hot spot strength are neglected $\sigma=0$ we obtain:
\beq
B_{\sigma=0}=\frac{\left(N_{q}-1\right)}{2\pi\left(B_p+2B_q\right){N_q}^2}.
\eeq
\subsection{Two  Different Hot Spots From Both Protons}
\par In the case II the two sums become one of two sub-cases. For $i_1\neq i_2$ but $j_1\neq j_2$ we get  for the sums :
\beq
\left(\sum_{i_{1},j_{1}}p_{i_1}\tilde{p}_{j_1}e^{-\frac{\left(\vec{a}_{i_{1}}-\vec{b}_{j_{1}}\right)^{2}}{4B_{q}}}\right)\left(\sum_{i_{2},j_{2}}p_{i_2}\tilde{p}_{j_2}e^{-\frac{\left(\vec{a}_{i_{2}}-\vec{b}_{j_{2}}\right)^{2}}{4B_{q}}}\right)\rightarrow p_{1}\tilde{p}_{1}p_{2}\tilde{p}_{2}N_q^2\left(N_q-1\right)^2e^{-\frac{\left(\vec{a}_1-\vec{b}_{1}\right)^{2}+\left(\vec{a}_2-\vec{b}_{2}\right)^{2}}{4B_{q}}},
\eeq
\par with a factor of $\left(I\left(\vec{0},\vec{s}_1\right)I\left(\vec{b},\vec{s}_2\right)\right)^{N_q-2}$ we get the integral to be:
\beq
C=\frac{p_{1}\tilde{p}_{1}p_{2}\tilde{p}_{2}\left(N_{q}-1\right)^{2}}{4N_{q^{4}}\left(2\pi\right)^{8}B_{p}^{2}B_{q}^{2}}\left(\int d^{7}xe^{-\vec{x}^{T}M_C\vec{x}}\right)^{2},
\eeq
\par with the vectors being $\vec{x}^{T}=\left(a_{1x},a_{2x},b_{1x},b_{2x},b_x,s_{1x},s_{2x}\right)$ and the matrix:
\beq
M_C=\left(\begin{array}{ccccccc}
\frac{B_{p}+2B_{q}}{4B_{p}B_{q}} & 0 & -\frac{1}{4B_{q}} & 0 & 0 & -\frac{i}{2N_{q}} & 0\\
0 & \frac{B_{p}+2B_{q}}{4B_{p}B_{q}} & 0 & -\frac{1}{4B_{q}} & 0 & -\frac{i}{2N_{q}} & 0\\
-\frac{1}{4B_{q}} & 0 & \frac{B_{p}+2B_{q}}{4B_{p}B_{q}} & 0 & -\frac{1}{2B_{p}} & 0 & -\frac{i}{2N_{q}}\\
0 & -\frac{1}{4B_{q}} & 0 & \frac{B_{p}+2B_{q}}{4B_{p}B_{q}} & -\frac{1}{2B_{p}} & 0 & -\frac{i}{2N_{q}}\\
0 & 0 & -\frac{1}{2B_{p}} & -\frac{1}{2B_{p}} & \frac{1}{B_{p}} & 0 & \frac{i}{N_{q}}\\
-\frac{i}{2N_{q}} & -\frac{i}{2N_{q}} & 0 & 0 & 0 & \frac{N_{q}-2}{2N_{q}^{2}}B_{p} & 0\\
0 & 0 & -\frac{i}{2N_{q}} & -\frac{i}{2N_{q}} & \frac{i}{N_{q}} & 0 & \frac{N_{q}-2}{2N_{q}^{2}}B_{p}
\end{array}\right).
\eeq
\par Using the Gaussian formula we get:
\beq
C=\frac{p_{1}\tilde{p}_{1}p_{2}\tilde{p}_{2}\left(N_q-1\right)^2}{8\pi\left(B_p+B_q\right)N_q^2},
\eeq
\par average over the hot spots strength gives us the final form for this case:
\beq
C=\frac{E\left[p\right]^4\left(N_q-1\right)^2}{8\pi\left(B_p+B_q\right)N_q^2}=\frac{\left(N_q-1\right)^2e^{2\sigma^2}}{8\pi\left(B_p+B_q\right)N_q^2},
\label{caseIIIfinal}
\eeq
If the fluctuations of the hot spot strength are neglected, $\sigma=0$ and we obtain 
\beq
C_{\sigma=0}=\frac{\left(N_q-1\right)^2}{8\pi\left(B_p+B_q\right)N_q^2}.
\eeq
\section{Total Effective Cross section and conclusions.}
\par Putting together our results in Eq. \ref{pdfassig},  Eq. \ref{caseIfinal}, Eq. \ref{caseIIfinal}, and Eq. \ref{caseIIIfinal} we 
 find $\sigma_{eff}$ to be:
\beq
\sigma_{eff}=\left(A+B+C\right)^{-1}=\frac{8\pi N_q^2}{\frac{e^{2\sigma^2}}{B_q}+\frac{4\left(N_q-1\right)e^{\sigma^2}}{B_p+2B_q}+\frac{\left(N_q-1\right)^2}{B_p+B_q}}.
\eeq

Here we normalized the gluon density to one according Eq. \ref{norm1} using Eq.\ref{pdfassig}.
\par Using the parameters from table I in \cite{Mantysaari:2022ffw}, which for convenience we present here as Table 1,
 we get two possible values for $\sigma_{eff}$:
\begin{table}
\begin{tabular}{|c|c|c|c|}
\hline 
Parameter & Description & Variable $N_{q}$ & $N_{q}\equiv3$\tabularnewline
\hline 
\hline 
$N_{q}$ & Number of hot spots & $6.79_{-4.83}^{+2.93}$ & $3$\tabularnewline
\hline 
$\sigma$ & Magnitude of hot spots strength fluctuations & $0.833_{-0.441}^{+0.194}$ & $0.563_{-0.141}^{+0.143}$\tabularnewline
\hline 
$B_{q}\left[GeV^{-2}\right]$ & Hot spot size & $0.474_{-0.286}^{+0.434}$ & $0.346_{-0.202}^{+0.282}$\tabularnewline
\hline 
$B_{p}\left[GeV^{-2}\right]$ & Proton size & $4.02_{0.728}^{+1.73}$ & $4.45_{-0.803}^{+0.801}$\tabularnewline
\hline 
\end{tabular}
\caption{The four parameters used in our calculations as taken from \cite{Mantysaari:2022ffw} for the case of Variable $N_{q}$ and  $N_{q}\equiv3$}
\end{table}

\begin{eqnarray}
 \sigma_{eff}\approx17\,\,{\rm mb}\,\,\, {\rm for} N_q=7,\nonumber\\[10pt]
 \sigma_{eff}\approx10.5 {\rm mb}\,\, {\rm for} N_q=3.\nonumber\\[10pt]
 \label{result}
\end{eqnarray}

\par We see that the hot spot model with hot spot strength fluctuations taken into account, and especially for $N_q=3$ leads to
substantially larger cross section of the DPS process  than   obtained from experimental data \cite{ATLAS,CMS} (\cite{BS} for recent review).

 Moreover , note that $\sigma_{eff}\sim 20 $ mb is obtained for   virtualities of  $\sim \approx 20$ GeV and higher for hard processes. In order to obtain the values of $\sigma_{eff}$ at scales of order several GeV one needs to carry inverse DGLAP evolution leading to $\sigma_{eff}=$ 25-35 mb . The latter number is similar to the value of $\sigma_{eff}$ used in MC generators \cite{PYTHIA,HERWIG} for moderate $p_t$ of several GeV.
 Thus the inclusion of pQCD evolution only increases the tension.

\par If we look at the case with $\sigma=0$,i.e. neglect the  hot spot strength fluctuations we obtain that for the set with variable $N_{q}$ we find $\sigma_{eff}\approx32 mb$, and for the set with a set $N_{q}\equiv3$ we get $\sigma_{eff}\approx17 mb$.
Thus if we neglect  hot spot strength fluctuations, the DPS rate  is  consistent 
 with the experimental data, however it will be still  in tension  for  $N_q=3$ if we take into account the $Q^2$ pQCD evolution. On the other hand if we increase the hot spot number the DPS cross section  decreases and is consistent with the current DPS data. 
\par In conclusion we see that DPS information can be used as an effective constraint on the models of  the nucleon structure.
\acknowledgments The research of B. Blok and R. Segev was supported by ISF grant number 2025311 and BSF grant  2020115.
The research of M. Strikman was supported by BSF grant  2020115 and
by  US Department of Energy Office of Science, Office of Nuclear Physics under Award No. DE–FG02–93ER40771.


\begin{thebibliography}{}
\bibitem{EIC}R. Abdul Khalek et. al., Science Requirements and Detector Concepts for the Electron-Ion Collider: EIC Yellow Report, arXiv:2103.05419 [physics.ins-det].
\bibitem{BS}B.~Blok and M.~Strikman,
Adv. Ser. Direct. High Energy Phys. \textbf{29} (2018), 63-99.
[arXiv:1709.00334 [hep-ph]].
\bibitem{levin} E. M. Levin, L. L. Frankfurt, 
 UFN, 94 (1968), 243–288; Sov. Phys. Usp., 11 (1968), 106–129.
\bibitem{mueller1} A.~H.~Mueller,
Nucl. Phys. B Proc. Suppl. \textbf{18} (1991), 125-132.
\bibitem{hs1}H. Mäntysaari,  B. Schenke  
 Phys.Lett.B 772 (2017) 832-838  •  e-Print:  1703.09256 [hep-ph].
\bibitem{Mantysaari:2022ffw}
H.~M\"antysaari, B.~Schenke, C.~Shen and W.~Zhao,
Phys. Lett. B \textbf{833} (2022), 137348.

 \bibitem{Mantysaari:2022sux}
H.~M\"antysaari, F.~Salazar and B.~Schenke,
Phys. Rev. D \textbf{106} (2022) no.7, 074019.
[arXiv:2207.03712 [hep-ph]].
\bibitem{GW}M.~L.~Good and W.~D.~Walker,
Phys. Rev. \textbf{120} (1960), 1857-1860.

\bibitem{progressinphsyics}L.~Frankfurt, V.~Guzey, A.~Stasto and M.~Strikman,
Rept. Prog. Phys. \textbf{85} (2022) no.12, 126301
[arXiv:2203.12289 [hep-ph]].



\bibitem{TP}N.~Paver and D.~Treleani,
Z. Phys. C \textbf{28} (1985), 187.
\bibitem{M}M.~Mekhfi,
Phys. Rev. D \textbf{32} (1985), 2371.
\bibitem{16a}   J.R.\ Gaunt and W.J.\ Stirling,
	  JHEP {\bf 1003}, 005 (2010) ,[arXiv:0910.4347 [hep-ph]].
\bibitem{16b}
  B.\ Blok, Yu.\ Dokshitzer, L.\ Frankfurt and M.\ Strikman,
  Phys.\ Rev.\  D {\bf 83}, 071501 (2011)
  [arXiv:1009.2714 [hep-ph]].
 \bibitem{16c} M.~Diehl,
	  PoS D {\bf IS2010} (2010) 223
	  [arXiv:1007.5477 [hep-ph]].
\bibitem{16d} J.R.\ Gaunt and W.J.\ Stirling,
	  JHEP {\bf 1106},  048 (2011) [arXiv:1103.1888 [hep-ph]].	  
  \bibitem{16e} B.\ Blok, Yu.\ Dokshitser, L.\ Frankfurt and M.\ Strikman,
  Eur.\ Phys.\ J.\ C {\bf72}, 1963  (2012)
  [arXiv:1106.5533 [hep-ph]].
\bibitem{16g} M.\ Diehl, D.\ Ostermeier and A.\ Schafer,
	  JHEP {\bf 1203} (2012) 089
	  [arXiv:1111.0910 [hep-ph]].
 \bibitem{16k}
 B.~Blok, Y.~Dokshitzer, L.~Frankfurt and M.~Strikman,
  Eur.\ Phys.\ J.\ C {\bf 74} (2014) 2926
  [arXiv:1306.3763 [hep-ph]].
\bibitem{16l}
  M.~Diehl, J.~R.~Gaunt and K.~Schönwald,
  JHEP {\bf 1706} (2017) 083
  [arXiv:1702.06486 [hep-ph]].

\bibitem{16n}
  A.~V.~Manohar and W.~J.~Waalewijn,
  Phys.\ Rev.\ D {\bf 85} (2012) 114009.
  \bibitem{ATLAS}O.~Kuprash [ATLAS],
``Studies of the underlying-event properties and of hard double parton scattering with the ATLAS detector,''
PoS \textbf{DIS2017} (2018), 035.
\bibitem{CMS}R.~Gupta [CMS and TOTEM],
``Double parton scattering studies in CMS,''
PoS \textbf{EPS-HEP2021} (2022), 335.

\bibitem{diehl}M.~Diehl,
Phys. Rept. \textbf{388} (2003), 41-277
[arXiv:hep-ph/0307382 [hep-ph]].

\bibitem{radyushkin}A.~V.~Belitsky and A.~V.~Radyushkin,
Phys. Rept. \textbf{418} (2005), 1-387
[arXiv:hep-ph/050400.
\bibitem{PYTHIA}
T.~Sj\"ostrand,
Adv. Ser. Direct. High Energy Phys. \textbf{29} (2018), 191-225
[arXiv:1706.02166 [hep-ph]].
\bibitem{HERWIG}J.~Bellm, S.~Gieseke and P.~Kirchgaesser,
Eur. Phys. J. C \textbf{80} (2020) no.5, 469
[arXiv:1911.13149 [hep-ph]].

\end{thebibliography}
\end{document}